\documentclass[sigconf,nonacm]{acmart}

\usepackage{inputenc}
\usepackage[T1]{fontenc}
\usepackage{textcomp} 

\usepackage{amsmath,amsfonts}
\usepackage{algorithm}
\usepackage{algpseudocodex}
\usepackage{booktabs}
\usepackage{graphicx}
\usepackage[dvipsnames]{xcolor}
\usepackage{fancyhdr}

\usepackage{hyperref}
\hypersetup{
  colorlinks,
  citecolor=Violet,
  linkcolor=Red,
  urlcolor=Blue}

\usepackage{tabularx}

\usepackage{subcaption}  
\usepackage{caption}
\usepackage{wrapfig}

\usepackage{enumitem}
\setlist[itemize]{noitemsep, topsep=0pt, leftmargin=*}

\usepackage{xspace}

\usepackage[most]{tcolorbox}
\usepackage{listings}

\definecolor{codebg}{RGB}{245,245,245} 
\definecolor{codekeyword}{RGB}{0,0,255} 

\newtcblisting{mycode}[1]{
  arc=2pt,                          
  boxrule=0.5pt,                    
  colback=codebg,                   
  colframe=gray!50,                 
  listing only,
  listing options={
    style=tcblatex,
    language=#1,
    basicstyle=\ttfamily\small,
    keywordstyle=\color{codekeyword},
    breaklines=true,
    columns=fullflexible
  },
  sharp corners=downhill            
}

\usepackage{multirow}
\usepackage{threeparttable}

\usepackage[export]{adjustbox}

\usepackage{tikz}

\newcommand{\design}{HAVEN\xspace}

\usepackage{setspace}
\setlist[itemize]{noitemsep, topsep=0pt, leftmargin=*}

\AtBeginDocument{%
  }

\setcopyright{acmlicensed}
\copyrightyear{2018}
\acmYear{2018}
\acmDOI{XXXXXXX.XXXXXXX}
\acmConference[Conference acronym 'XX]{Make sure to enter the correct
  conference title from your rights confirmation email}{June 03--05,
  2018}{Woodstock, NY}
\acmISBN{978-1-4503-XXXX-X/2018/06}

\begin{document}

    \title{
        \design: \underline{H}igh-Bandwidth Flash \underline{A}ugmented \underline{V}ector \underline{EN}gine for Large-Scale Approximate Nearest-Neighbor Search Acceleration\\
    }


\author[P.-K. Hsu]{Po-Kai Hsu}
\authornote{Equal contribution}

\affiliation{
  \institution{Georgia Institute of Technology}
  \city{Atlanta}
  \country{United States}
}
\email{pokai.hsu@gatech.edu}

\author[W. Xu]{Weihong Xu}
\authornotemark[1]
\affiliation{
  \institution{Ecole Polytechnique Fédérale de Lausanne}
  \city{Lausanne}
  \country{Switzerland}
}
\email{weihong.xu@epfl.ch}

\author[Q. Liu]{Qunyou Liu}
\affiliation{
  \institution{Ecole Polytechnique Fédérale de Lausanne}
  \city{Lausanne}
  \country{Switzerland}
}
\email{qunyou.liu@epfl.ch}

\author[T. Rosing]{Tajana Rosing}
\affiliation{
  \institution{University of California, San Diego}
  \city{La Jolla}
  \country{United States}
}
\email{tajana@ucsd.edu}

\author[S. Yu]{Shimeng Yu}
\affiliation{
  \institution{Georgia Institute of Technology}
  \city{Atlanta}
  \country{United States}
}
\email{shimeng.yu@ece.gatech.edu}


\begin{abstract}
Retrieval-Augmented Generation (RAG) relies on large-scale Approximate Nearest Neighbor Search (ANNS) to retrieve semantically relevant context for large language models. Among ANNS methods, IVF-PQ offers an attractive balance between memory efficiency and search accuracy. However, achieving high recall requires reranking which fetches full-precision vectors for reranking, and the billion-scale vector databases need to reside in CPU DRAM or SSD due to the limited capacity of GPU HBM. This off-GPU data movement introduces substantial latency and throughput degradation.

We propose HAVEN, a GPU architecture augmented with High-Bandwidth Flash (HBF) which is a recently introduced die-stacked 3D NAND technology engineered to deliver terabyte-scale capacity and hundreds of GB/s read bandwidth. By integrating HBF and near-storage search unit as an on-package complement to HBM, HAVEN enables the full-precision vector database to reside entirely on-device, eliminating PCIe and DDR bottlenecks during reranking.

Through detailed modeling of re-architected 3D NAND subarrays, power-constrained HBF bandwidth, and end-to-end IVF-PQ pipelines, we demonstrate that HAVEN improves reranking throughput by up to 20$\times$ and latency up to 40$\times$ across billion-scale datasets compared to GPU-DRAM and GPU-SSD systems. Our results show that HBF-augmented GPUs enable high-recall retrieval at throughput previously achievable only without reranking, offering a promising direction for memory-centric AI accelerators.
\end{abstract}





\maketitle

\section{Introduction}

\begin{figure}[t]
    \centering
    \includegraphics[width=\linewidth]{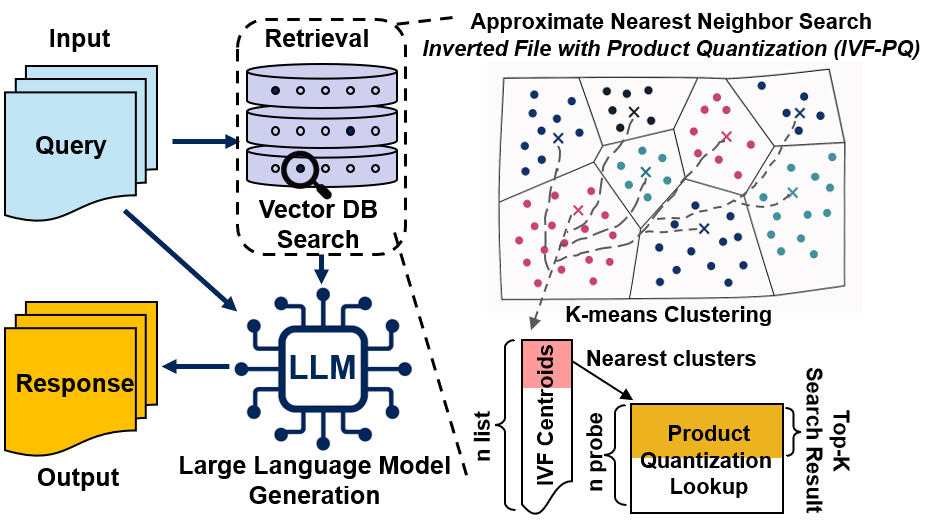}
    \caption{Overview of the retrieval-augmented generation (RAG) pipeline. Retrieval of the relevant knowledge from vector database relies on approximate nearest neighbor search.}
    \label{fig:RAG_Intro}
\end{figure}

Retrieval-Augmented Generation (RAG) has become a core technique for large-scale language models by combining parametric generation with non-parametric external knowledge sources. A crucial component in RAG is the retrieval module, which searches extremely large vector databases to return semantically relevant embeddings as illustrated in Figure ~\ref{fig:RAG_Intro}. Approximate nearest neighbor search (ANNS) algorithms are therefore essential for enabling low-latency retrieval at billion-scale.

Among ANNS algorithms, the inverted file with product quantization (IVF-PQ) framework has proven highly effective in balancing accuracy, memory footprint, and search efficiency for large-scale vector databases~\cite{jegou2011product, pan2020pqdual, tian2024billionann}. IVF-PQ first partitions the search space via coarse quantization and then applies product quantization (PQ) to compress the stored vectors. To improve recall, modern IVF-PQ implementations perform a \emph{reranking} stage, where the original full-dimensional vector is fetched and re-ranked~\cite{yang2024fastann}. However, this  reranking step requires access to the uncompressed vector database, whose size often reaches hundreds of gigabytes or more at billion-vector scale~\cite{tian2024billionann}.

Current GPUs integrate high-bandwidth memory (HBM) within the device package, providing exceptional bandwidth but limited capacity (typically 80--192~GB per device)~\cite{HBM}. As a result, the full IVF-PQ database cannot fit in HBM, forcing systems to store the original vectors in host DDR memory or NVMe SSDs. The resulting off-GPU data movement introduces orders-of-magnitude higher latency and significantly degrades ANNS throughput, particularly during the reranking stage.

To overcome this capacity bottleneck, emerging research and industry roadmaps propose High-Bandwidth Flash (HBF), a die-stacked NAND flash integrated in a manner analogous to HBM~\cite{sandisk2025hbf}. HBF can potentially deliver 8--16$\times$ higher capacity per stack while maintaining hundreds of GB/s read bandwidth through massively parallel 3D NAND subarrays. Such characteristics make HBF an attractive candidate for GPU co-packaged memory in read-intensive workloads.

In this work, we introduce \textbf{HAVEN}, an architecture that replaces a portion of HBM capacity with HBF inside the GPU package. On top of HBF, we also integrate near-storage search unit to further improve the reranking performance. With HAVEN, the complete vector database including full-precision vectors required for IVF-PQ reranking can be stored and processed on-package, eliminating DDR/SSD round-trips and significantly reducing retrieval latency. We analyze bandwidth and latency trade-offs, study architectural integration constraints, and evaluate the performance impact on large-scale ANNS pipelines representative of RAG systems. 

The main contributions of this paper are:
\begin{itemize}
    \item We quantify the memory capacity and bandwidth limitations of IVF-PQ on current HBM-only GPUs.
    \item We propose HAVEN, a hybrid HBM plus HBF GPU memory hierarchy with near-storage search unit enabling on-package storage and processing of massive vector databases.
    \item We model and evaluate IVF-PQ reranking performance on HAVEN, demonstrating up to 20$\times$ throughput and 40$\times$ latency improvements over DDR/SSD-backed systems.
\end{itemize}

\section{Background}

\subsection{Approximate Nearest Neighbor Search}
The goal of $k$-nearest neighbor ($k$-NN) search is to retrieve the $k$ points in a dataset $\mathcal{X}=\{x_1,\ldots,x_N\}$ that are closest to a query vector $q \in \mathbb{R}^D$. The exact search problem can be formulated as $\mathcal{R} = \underset{\mathcal{R} \subseteq \mathcal{X},\,|\mathcal{R}| = k}{\arg\min}\; dist(q, x)$, where $dist(\cdot,\cdot)$ may be Euclidean distance, cosine similarity, or inner product. A brute-force solution requires $\mathcal{O}(ND)$ comparisons, which becomes impractical for million- or billion-scale datasets. To meet real-time latency requirements, modern systems~\cite{johnson2019billion,hnsw} rely on approximate nearest neighbor search (ANNS), which returns an approximate neighbor set $\hat{\mathcal{R}}$ while examining only a small subset of the data. The performance of ANNS methods is largely determined by the index structure used to narrow the search space. Popular indices include inverted file (IVF) systems~\cite{johnson2019billion,ivf}, graph-based methods~\cite{diskann,hnsw}, and hashing-based approaches~\cite{he2013kmean_hash}, each offering different trade-offs between latency, memory, and accuracy.

\begin{lstlisting}[
    language=Python,
    basicstyle=\ttfamily\footnotesize,
    caption={\small Query search in IVF-PQ with reranking.},
    label={code:ivf_pq_query},
    float=t
  ]
Input: Query q, (Nprobe, Nrerank, k)
IVF-PQ Index (coarse_centroids, Nlists, PQ_codebooks)
# Step 1 Coarse Quantizer Probing
for each coarse centroid c_j:
    d_j = distance(q, c_j)
probe_IDs = indices of Nprobe smallest d_j
# Step 2 IVF-PQ Scan on Probed Lists
candidates = empty max-heap of size Nrerank
for each list_id in probe_IDs:
    for each vector code in list[list_id]:
        dist = PQ_distance(q, code)
        push (dist, vector_id) into candidates 
# Step 3: Rerank on top-k Candidates
for each (dist, id) in candidates:
    dist_exact[id] = exact_distance(q, base_vectors[id])
results = sort candidate ids by dist_exact ascending
return top-k results
\end{lstlisting}

\noindent
\textbf{IVF-PQ Search.}
In this work, we focus on the widely used IVF-PQ (Inverted File with Product Quantization) indexing technique, offering a good balance of search speed and memory efficiency for large-scale ANNS. As illustrated in Figure~\ref{fig:RAG_Intro}, IVF-PQ integrates naturally with LLM-based RAG pipelines. During offline index building, IVF applies $k$-means to partition the vector space into $N_{\text{list}}$ coarse clusters, while Product Quantization (PQ) further compresses the residual vectors within each cluster into short codes. This combination enables storing billions of vectors in memory while maintaining competitive retrieval accuracy. 
At query time, IVF-PQ performs three main steps (Listing~\ref{code:ivf_pq_query}):
1) \emph{coarse quantizer probing},  
2) \emph{approximate IVF-PQ scan} on the selected inverted lists, and  
3) \emph{reranking} the top-$k$ candidates using higher-precision distances.

\noindent
\textbf{Step 1: Coarse Quantizer Probing.}
Given a query $q$, IVF first computes its distances to all coarse centroids and selects the $N_{\text{probe}}$ closest ones. Only the corresponding inverted lists are searched, reducing the search space dramatically from $N$ vectors to a small fraction proportional to $N_{\text{probe}}$.

\noindent
\textbf{Step 2: IVF-PQ Scan on Probed Lists.}
Within each probed list, vectors are stored as compressed PQ codes. The approximate PQ distance between the query and compressed vectors can be computed via a lightweight table-lookup-and-sum procedure. This approximate scanning stage is extremely efficient and provides a coarse ranking of candidate neighbors.

\begin{figure}[t]
    \centering
    \begin{subfigure}[b]{0.9\linewidth}
        \centering
        \includegraphics[width=\linewidth]{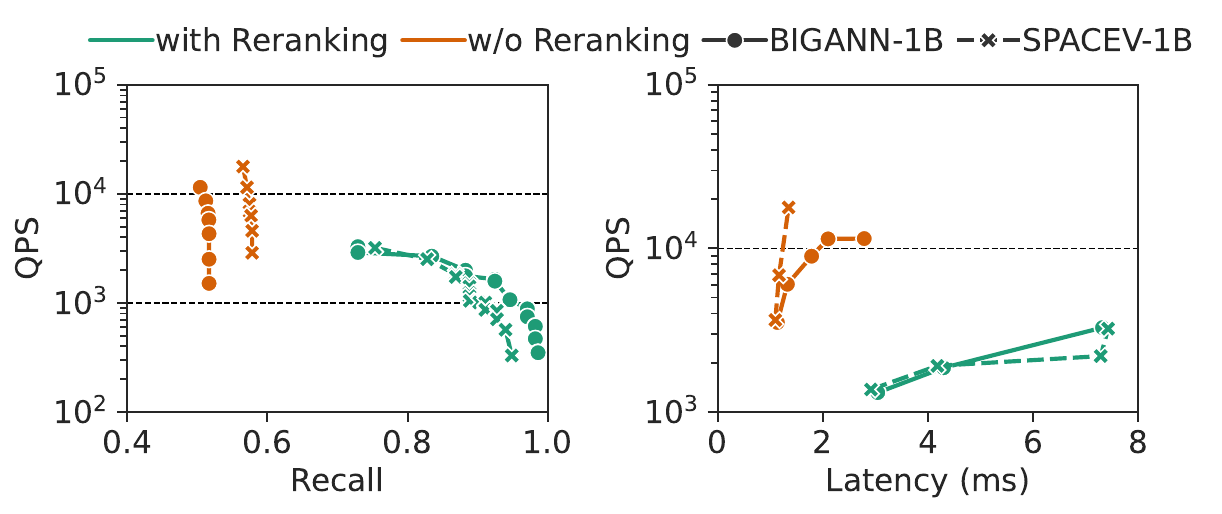}
        \caption{}
    \end{subfigure}   
    \hfill
    \begin{subfigure}[b]{0.45\linewidth}
        \centering
        \includegraphics[width=\linewidth]{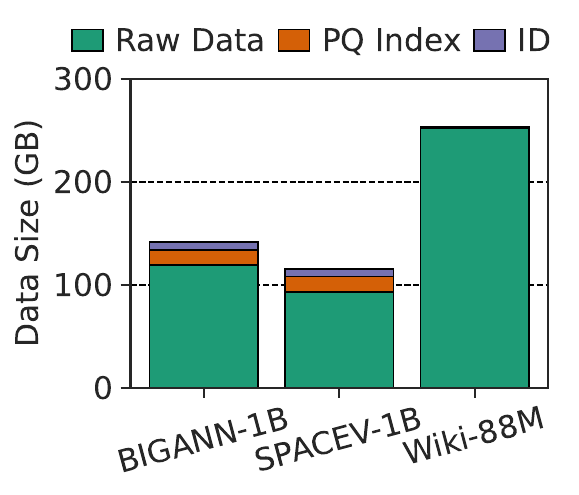}
        \caption{}
    \end{subfigure}   
    \begin{subfigure}[b]{0.45\linewidth}
        \centering
        \includegraphics[width=\linewidth]{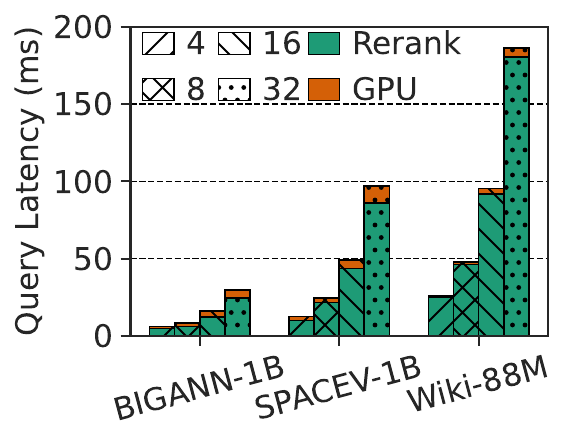}
        \caption{}
    \end{subfigure}   
    \caption{(a) Trade-offs between throughput, recall, and latency with re-rank, (b) Memory consumption breakdown, and (c) Query time breakdown.}
    \label{fig:profiling}
\end{figure}

\noindent
\textbf{Step 3: Reranking on top-$k$ Candidates.}
Since PQ codes introduce quantization error, IVF-PQ reranks the top-$k$ candidates by recomputing distances using full-precision vectors. This reranking operates on a small set of $N_{\text{rerank}}$ vectors and significantly boost recall as shown in Figure~\ref{fig:profiling}(a), making IVF-PQ both scalable and accurate for high-quality retrieval.

\subsection{Profiling IVF-PQ-based ANNS}
GPUs are widely used for ANNS because of their high bandwidth and parallelism, but billion-scale datasets quickly exceed the limited capacity of GPU's HBM. As a result, systems typically store only the IVF index and PQ codes on the GPU, offloading full-precision vectors for reranking to host DRAM or SSD. To quantify the impact of this design, we profile a system that our lab had access to, which is equipped with DDR4 memory, NVMe SSD, and an NVIDIA A100 GPU, as described in Section~\ref{subsec:methodology}. We observe the following limitations:

\noindent\textbf{Challenge 1: Imbalance between throughput, recall, and latency.}
Figure~\ref{fig:profiling}(a) shows that enabling reranking substantially improves recall but sharply reduces throughput and increases query latency. Without reranking, the system achieves higher QPS but suffer degraded recall; with reranking, recall increases but throughput collapses by one to two orders of magnitude. This imbalance makes it difficult to simultaneously satisfy high-recall and low-latency requirements, especially under real-time or batch-processing constraints.

\noindent\textbf{Challenge 2: Memory footprint challenge.}
As shown in Figure~\ref{fig:profiling}(b), raw vectors consume the overwhelming majority of memory capacity—hundreds of GBs for datasets such as Wiki-88M and SIFT-1B. In billion-scale deployments, storing both PQ codes on the GPU and raw vectors for reranking on CPU memory or SSD greatly inflates total memory demand. Limited GPU memory forces frequent data transfers, and DRAM/SSD capacity constraints make scaling beyond hundreds of millions of vectors costly or infeasible.

\noindent\textbf{Challenge 3: Long re-rank latency.}
As shown in Figure~\ref{fig:profiling}(c), reranking dominates query time because full-precision vectors must be fetched from DRAM or SSD across PCIe, adding substantial latency. This overhead grows at billion scale, especially with larger batches. Since raw vectors consume most of the memory footprint (Figure~\ref{fig:profiling}(b)), they cannot reside on the GPU, making off-device access unavoidable and leaving reranking fundamentally constrained by memory capacity and cross-device data movement.

\subsection{Novel High-Bandwidth Flash}
The bottlenecks described above stem from a fundamental mismatch between the memory requirements of IVF-PQ reranking and the limited capacity of on-package HBM. A100-class GPU provides excellent bandwidth but only a few tens of gigabytes of capacity, which is far short of storing PQ codes together with the full-precision vectors. Recent development in 3D non-volatile memory have introduced High Bandwidth Flash (HBF)~\cite{sandisk2025hbf}, a die-stacked NAND architecture designed to resemble HBM in form factor and signaling. Unlike SSDs that are optimized for density, HBF exposes wide internal parallelism and finer access granularity. Industry projections~\cite{sandisk2025hbf} suggest that HBF can deliver eight to sixteen times the capacity of HBM within a similar footprint while sustaining hundreds of GB per second of read bandwidth, which is well suited for read-intensive vector retrieval workloads.

\begin{figure}[t]
    \centering
    \includegraphics[width=\linewidth]{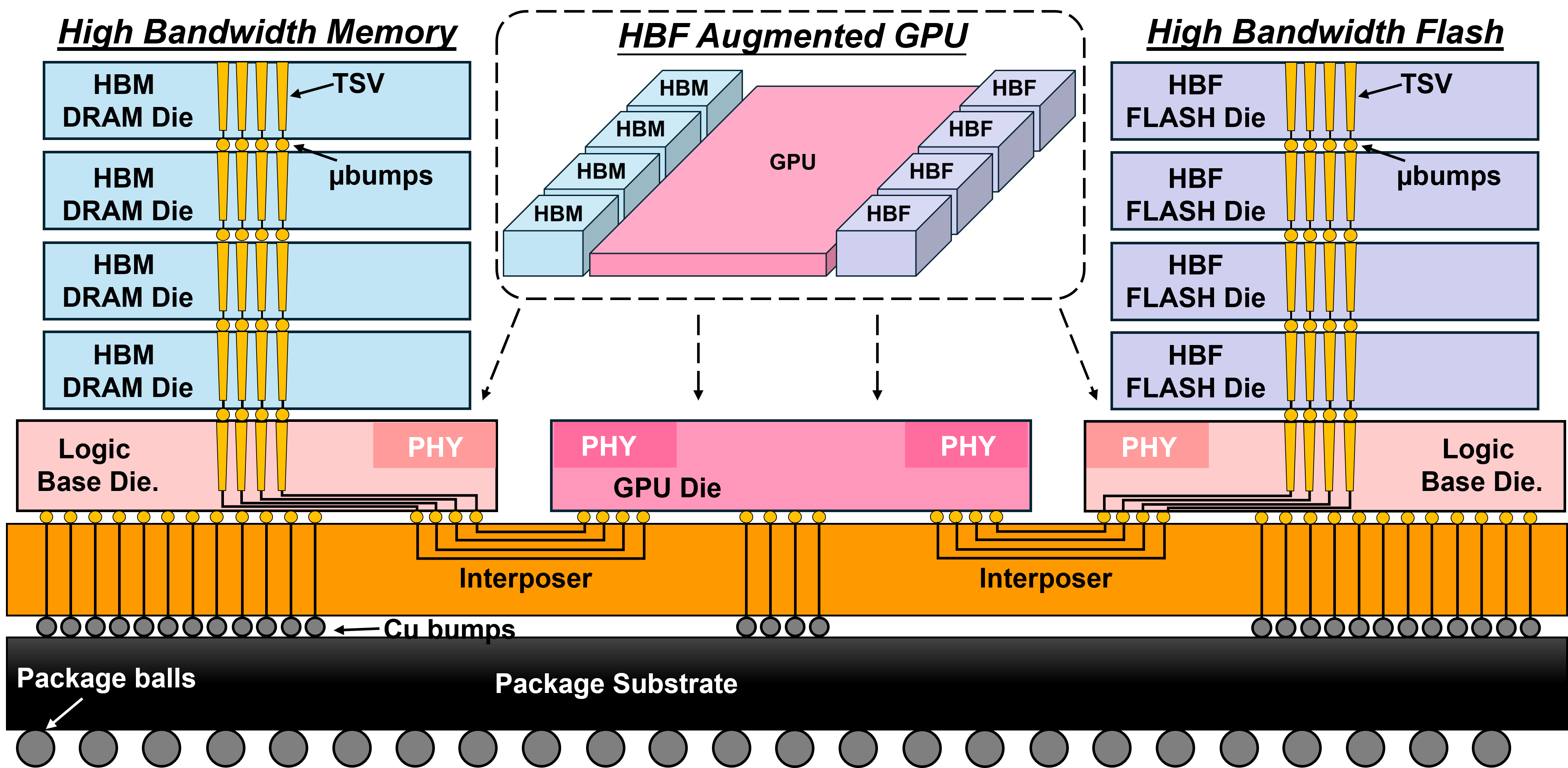}
    \caption{Overview of the HBF-augmented GPU. HBF stacks replace part of the HBM stacks and are co-packaged with GPU through 2.5D interposer.}
    \label{fig:HBF_Intro}
\end{figure}

\noindent\textbf{HBF-Augmented GPU.}
When HBF replaces part of the HBM stack inside the GPU package, the memory hierarchy gains an on-device, high-capacity tier as illustrated in Figure ~\ref{fig:HBF_Intro}. Storing full-precision vectors directly in HBF removes the PCIe and DDR bottlenecks that dominate reranking. This integration yields three direct benefits: (1) \textbf{lower latency} since reranking traffic no longer leaves the package, (2) \textbf{sufficient capacity} to store billion-scale raw vectors entirely on the device, and (3) \textbf{high reranking bandwidth} enabled by the extensive parallelism of distributed NAND subarrays.

\noindent\textbf{Impact on IVF-PQ reranking.}
With raw vectors kept in-package, reranking shifts from a latency-limited stage into a high-throughput operation. As a result, high-accuracy reranking becomes feasible without the severe throughput collapse seen in CPU backed or SSD backed systems. Larger batches become practical as well, enabling higher QPS in both real-time and offline RAG pipelines. In summary, HBF shifts the design toward a memory-centric approach that expands effective GPU capacity and keeps reranking on the device. Building on this opportunity, we propose HAVEN, a hybrid HBM plus HBF vector search engine that stores and processes raw vectors in HBF while reserving HBM for latency critical PQ operations. The next section describes the overall architecture along with the underlying 3D NAND re-architecture and hardware design.

\section{Proposed \design Accelerator}

\subsection{Architecture Overview}

\begin{figure}[t]
    \centering
    \includegraphics[width=\linewidth]{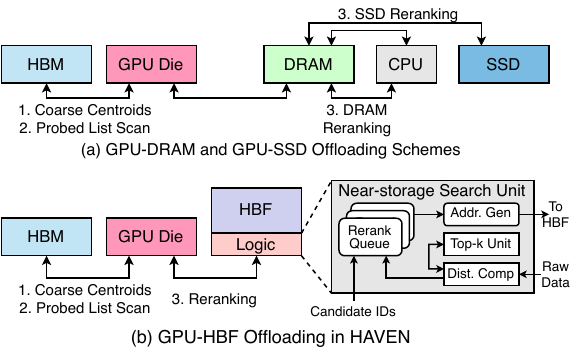}
    \caption{Dataflow comparison of (a) GPU-DRAM, GPU-SSD, and (b) \design's GPU-HBF offloading schemes.}
    \label{fig:dataflow}
\end{figure}

\noindent\textbf{Dataflow.}
\design adopts an efficient datapath by co-packaging HBF with the GPU die and HBM, as shown in Figure~\ref{fig:HBF_Intro}. The first two IVF-PQ stages (coarse centroid probing and IVF-PQ list scanning) remain on the GPU and HBM to preserve low-latency processing, while the final storage-intensive reranking stage is offloaded to HBF with near-storage processing (NSP). This design significantly reduces reranking latency compared to DRAM- or SSD-based systems. Figure~\ref{fig:dataflow} highlights the difference: in conventional designs (Figure~\ref{fig:dataflow}(a)), full-precision vectors must be fetched from CPU memory or SSD, incurring costly DDR or PCIe transfers. In contrast, \design’s GPU–HBF dataflow (Figure~\ref{fig:dataflow}(b)) keeps reranking entirely on-package, eliminating off-chip round-trips and leveraging HBF’s high internal bandwidth and parallelism.

\noindent\textbf{Near-Storage Search Unit.}
Because reranking has low arithmetic intensity, transferring raw vectors from HBF to HBM would introduce unnecessary data movement. To avoid this inefficiency and reduce query latency, \design performs reranking directly near HBF using a near-storage search unit implemented on the logic die beneath the HBF stack. As shown in Figure~\ref{fig:dataflow}(b), the unit first receives candidate IDs from the GPU and stores them in the \textbf{Rerank Queue}. \design includes $32\times$ Rerank Queues, each with a 8~KB buffer capable of storing up to 1,024 32-bit IDs and their distances. Each ID is then passed to the \textbf{Address Generation} module, which computes the physical address of its corresponding raw vector. The generated address is packaged into a read request and issued to HBF. Once the full-precision vector is fetched, the \textbf{Distance Computation Module}, equipped with 32 multiply–accumulators (MACs), computes its exact distance to the query. Finally, the \textbf{Top-$\mathbf{k}$ Unit}, implemented with a parallel Bitonic sorter, selects the best-matching candidates. These components enable fully on-package reranking, minimizing latency and maximizing the bandwidth benefits of HBF.

\subsection{3D NAND Re-Architecture for HBF}
Adapting NAND technology for HBF requires rethinking the structure of commercial 3D NAND. Commercial NAND is engineered primarily for density rather than bandwidth. The performance do not match the fine-grained and highly parallel access patterns needed for large-scale IVF-PQ reranking. To address this gap, we adopt a distributed subarray organization that increases internal parallelism and improves aggregate throughput. This section describes the re-architecture process, the subarray modeling framework, and the full-stack HBF modeling flow.

\begin{figure}[t]
    \centering
    \includegraphics[width=\linewidth]{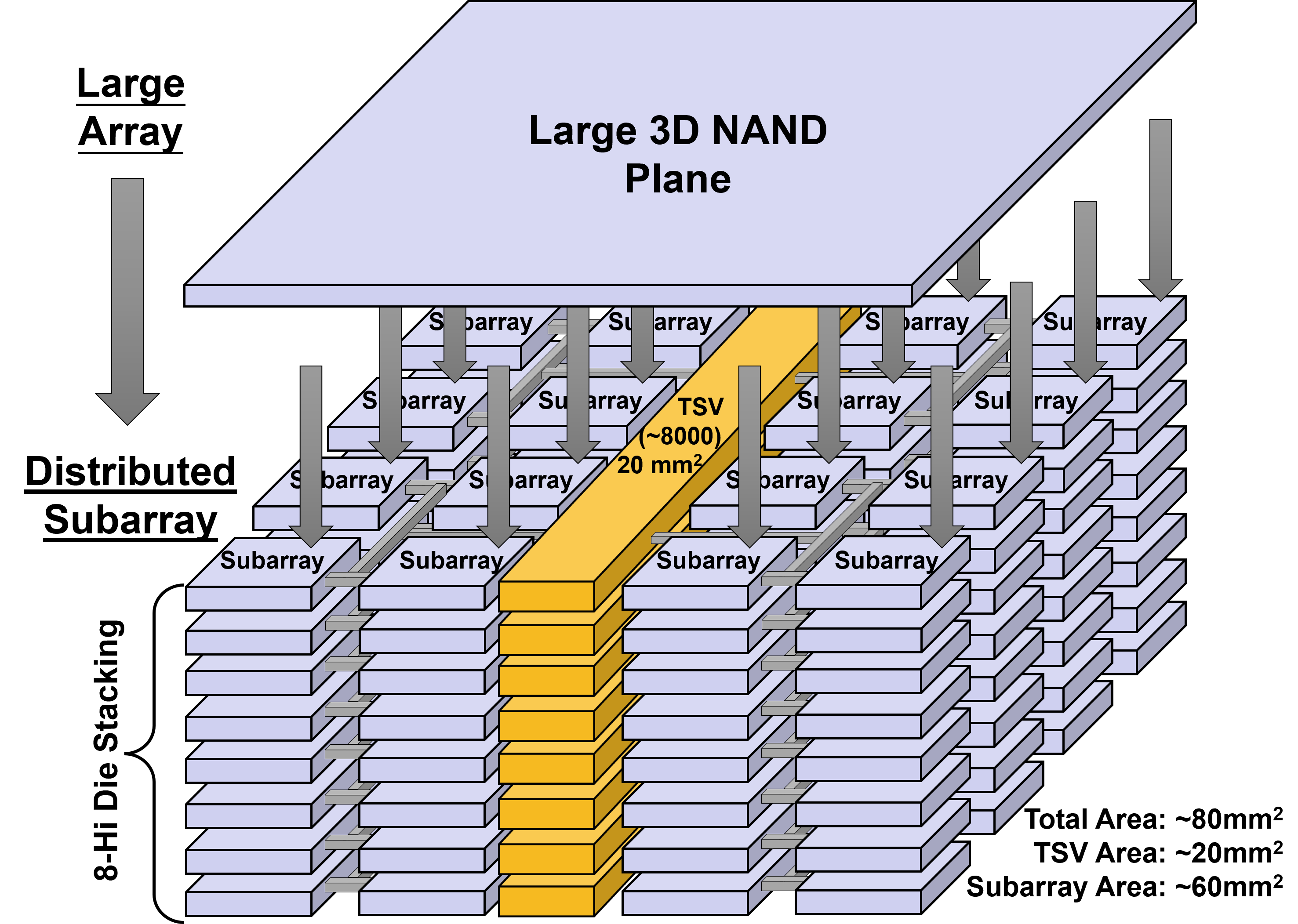}
    \caption{3D NAND re-architecture for HBF. The gigantic 3D NAND plane is re-architected to distributed subarrays for higher parallelism and higher bandwidth for HBF. The TSV occupies the center region.}
    \label{fig:HBF_3DNAND_Rearchitect}
\end{figure}

\noindent\textbf{3D NAND Re-Architecture.} Conventional 3D NAND is built around gigantic monolithic memory planes optimized for density rather than bandwidth. Such large arrays exhibit long bitline lengths, high wordline parasitic capacitance, and coarse access granularity. However, these characteristics do not align with the fine-grained access patterns required for IVF-PQ reranking. To address this mismatch, we move from monolithic planes to distributed small subarrays as illustrated in Figure ~\ref{fig:HBF_3DNAND_Rearchitect} similar to fast NAND approach~\cite{SamsungZNAND}. This architectural shift shortens local bitlines, reduces parasitics, improves access energy, and exposes significantly more parallel subarrays.

\begin{figure}[t]
    \centering
    \includegraphics[width=0.8\linewidth]{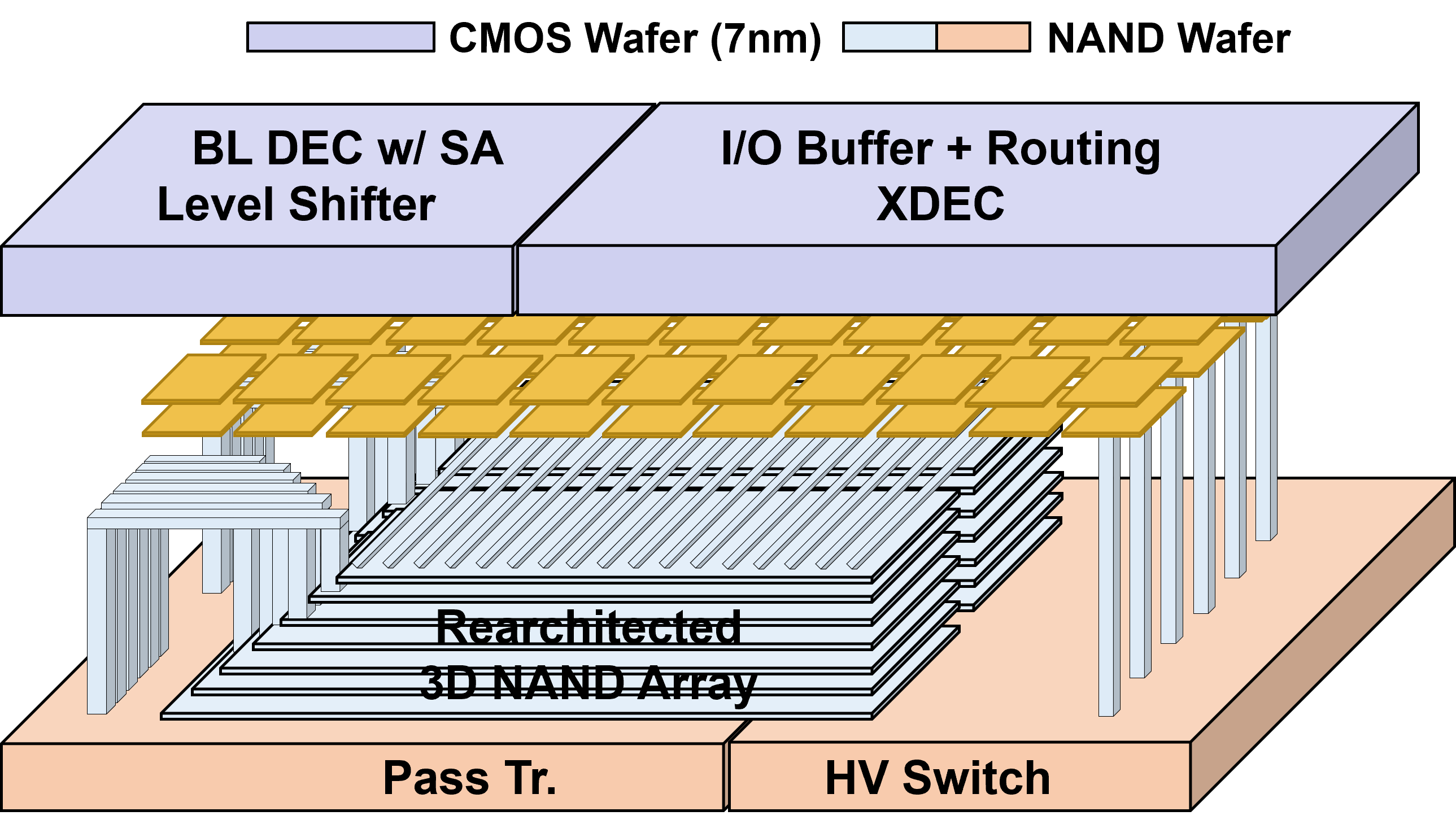}
    \caption{Heterogeneous integration for compact 3D NAND subarray form factor. CMOS-under-array integrate high voltage circuits including wordline and string select line drivers. Low voltage circuit including page buffer and I/O buffer are deployed on top advanced CMOS tier in 7nm technology node bonded with 3D NAND tier.}
    \label{fig:HeterogeneousIntegration}
\end{figure}

\noindent\textbf{Distributed Subarray Modeling.} We evaluate the architectural tradeoffs of distributing 3D NAND into a large number of small subarrays. In this architecture, each subarray operates with its own local wordline/bitline drivers and sense-amplifiers, enabling higher internal parallelism than conventional NAND organizations. We also incorporate heterogeneous integration~\cite{Wonbo3DNAND} to further increase memory density and improve performance as shown in Figure ~\ref{fig:HeterogeneousIntegration}.

\begin{figure}[t]
    \centering
    \includegraphics[width=0.8\linewidth]{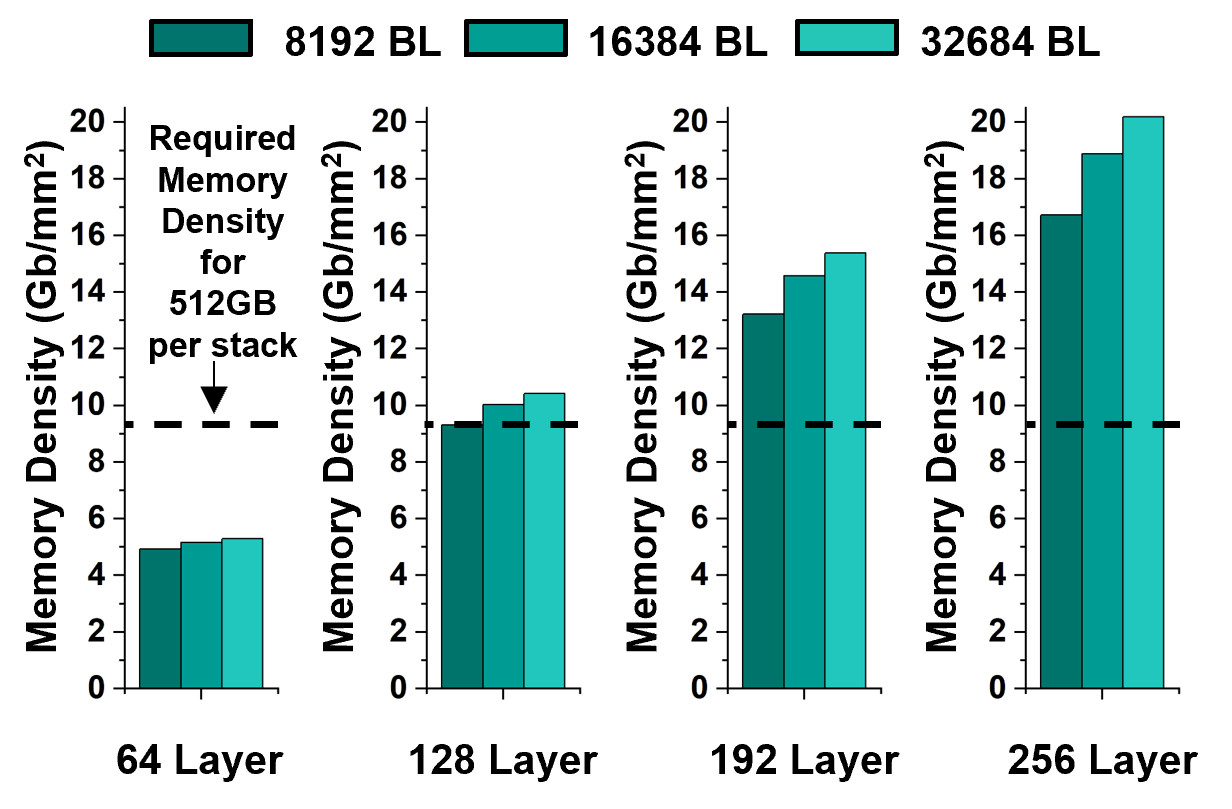}
    \caption{Simulation result of memory density versus number of bitline and number of wordline layers.}
    \label{fig:CapacityVsNOofLayer}
\end{figure}

To quantify the implications of this restructuring, we employ a distributed 3D NAND subarray modeling framework based on 3D-FPIM~\cite{3DFPIM} combined with NeuroSim~\cite{NeuroSim} for scaled peripheral circuits. This combination enables array-to-die-level projections for subarray granularity across parameters such as number of 3D layers, page size and block size. The detailed 3D NAND parameters are listed in Table ~\ref{table:sim_para}.
Using this simulation framework, we first sweep the number of WL layers in the 3D stack and evaluate density scaling trends as shown in Figure ~\ref{fig:CapacityVsNOofLayer}. The analysis shows that achieving 512 GB in an 8-Hi stack~\cite{sandisk2025hbf} requires at least 128 WL layers. To maintain design margin for more vector databases and align with the state-of-the-art NAND technology roadmaps~\cite{Samsung3DNAND280L, Samsung3DNAND4XXL}, we adopt 256 WL layers as our baseline technology node for all subsequent HBF modeling.

\begin{table}[t]
    \centering
    \footnotesize
    \caption{3D NAND Subarray Simulation Parameters}
    \begin{tabular}{|c|c|c|}
        \hline
        \textbf{} & \textbf{Parameters} & \textbf{Values} \\
        \hline
        \textbf{Advanced} & Technology & 7 nm FinFET Process \\
        \textbf{CMOS Tier} & VDD1 & 0.7 V \\
        \hline
        \textbf{3D NAND} & VC Hole Diameter & 145 nm \\
        \textbf{Physical} & BL Pitch & 40 nm \\
        \textbf{Parameter~\cite{3DFPIM}} & VC Hole Pitch & 248 nm \\
        \textbf{} & No. of WL & 64/128/196/256 \\
        \textbf{} & No. of SSL & 2 \\
        \textbf{} & No. of Sub Block & 2 \\
        \textbf{} & No. of BL & 1/2/4 KB \\
        \textbf{} & No. of Block & 64/128/256/512/1024 \\
        \textbf{} & No. of Bit & 1 (SLC) \\
        \textbf{} & WL Staircase Pitch & 725 nm \\
        \hline
        \textbf{CMOS} & Technology & 22 nm Process \\
        \textbf{under Array} & VDD2 & 0.9V \\
        \hline
        \end{tabular}
    \label{table:sim_para}
\end{table}

\noindent\textbf{High-Bandwidth Flash Modeling.} We integrate the distributed subarray model into a full-stack HBF modeling. Following the HBM stacking organization, we assume the reasonable TSV pitch for microbump of 50$~\mu$m~\cite{HBMTSVPitch} and maintain the TSV organization~\cite{skhynix2022hbm3demonstration} and an 8-Hi die-stack configuration~\cite{HBM} in Figure ~\ref{fig:HBF_3DNAND_Rearchitect} to enable fair comparison in capacity, energy, and throughput scaling. Note that we align with HBM2E specification integrated in A100 GPU.

\begin{figure}[t]
    \centering
    \includegraphics[width=\linewidth]{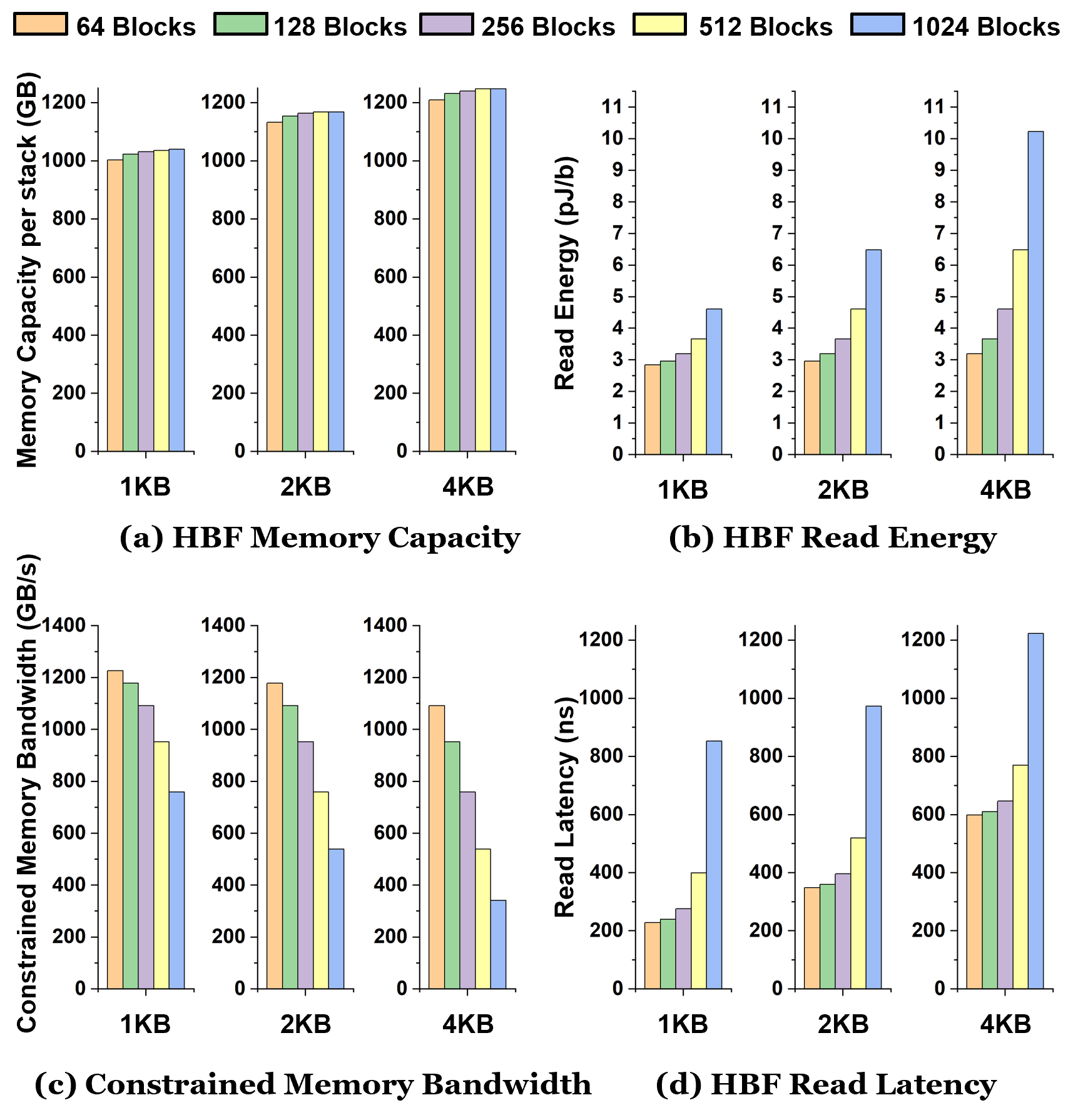}
    \caption{Simulation results of 8-Hi HBF configurations of page size = 1KB, 2KB and 4KB and block size = 64, 128, 256, 512 and 1024. (a) HBF memory capacity. (b) HBF read energy. (c) Power constrained memory bandwidth. (d) HBF read latency.}
    \label{fig:3DFPIM_Results}
\end{figure}

Simulation results based on 3D-FPIM subarray modeling reveal the expected overall trend as shown in Figure ~\ref{fig:3DFPIM_Results}. Larger subarrays (i.e., larger page sizes or more blocks per plane) provide higher capacity density as in Figure ~\ref{fig:3DFPIM_Results}(a). However, they also raise read energy and read latency as shown in Figures ~\ref{fig:3DFPIM_Results}(b) and ~\ref{fig:3DFPIM_Results}(d) due to longer intra-array wires and increased 3D structural parasitics. Conversely, aggressively shrinking subarrays reduces read energy and improves per-access latency but reduces overall capacity.

For energy consumption, we calibrate read energy using industry-grade 3D NAND data, scaling from the $\sim$30 pJ/bit read energy reported for contemporary devices~\cite{Kioxia_IMW}, and also account for HBM2 data-movement energy~\cite{oconnor2017finegrained}. Across distributed subarray configurations, read energy increases rapidly with larger page sizes or higher block counts, eventually becoming a dominant constraint on achievable bandwidth.

Due to fine-grained parallel distributed subarray architecture, bandwidth in HBF is fundamentally constrained by power delivery. To preserve thermal and package constraints comparable to HBM, we adopt an HBM-like 30 W power envelope per stack~\cite{son2023thermal}. Using this constraint, we sweep page sizes and block counts to estimate sustainable read bandwidth as shown in Figure ~\ref{fig:3DFPIM_Results}(d). As capacity increases from larger pages and more blocks, the power required per access grows and eventually limits the effective throughput. The resulting bandwidth degradation becomes especially severe when block count exceeds 128. To be compatible with HBM2E infrastructure, we assume the maximum bandwidth of 460 GB/s per stack~\cite{skhynix2019hbm2e}.

Read latency is also affected as shown in Figure ~\ref{fig:3DFPIM_Results}(d). As we scale subarray dimensions, the array RC delay and page activation time increase, leading to latency growth across configurations. Modeling results confirm that both capacity and performance scale in opposing directions, and no single design maximizes both.

Considering these tradeoffs, we select a configuration that provides the needed capacity while preserving bandwidth and keeping moderate latency. We select a 4 KB page size, 64 blocks per subarray, and 256 WL layers as the baseline HBF configuration for the remainder of our architectural analysis. This configuration provides high parallelism, manageable read energy within the 30 W envelope, and adequate capacity to store billion-scale vectors on-package for IVF-PQ reranking.

\section{Evaluation}\label{sec:eval}

\subsection{Methodology}\label{subsec:methodology}
\noindent\textbf{\design and HBF Simulation.} 
We simulate the re-architected 3D NAND subarray using 3D-FPIM~\cite{3DFPIM} combining with NeuroSim~\cite{NeuroSim}. The energy model is calibrated by industry-grade performance~\cite{Kioxia_IMW}. The ASIC near-storage search unit is implemented using Verilog HDL and synthesized with 22~nm technology node. Search unit achieves a clock frequency of 1GHz. 
Table~\ref{tab:hard_imp} summarizes the area and power results of the near-storage search unit, showing a total footprint of 4.11~mm$^2$ and 620.3~mW, with rerank queues dominating area and the Bitonic sorter dominating power.

\begin{table}[ht]
    \caption{Area and Power Results of Near-storage Search Unit}
    \centering
    \scriptsize
    \renewcommand{\arraystretch}{1.05}
    \begin{tabular}{c|c|c|c}
        \hline
        \textbf{Module} & \textbf{Size}  & \textbf{Area (mm$^2$)}  & \textbf{Power (mW)} \\
        \hline
        \hline
        \textbf{Rerank Queues}  & $\times 32$ & 3.84 & 122.6 \\
        \hline
        \textbf{Distance Computation Module} & $32\times$ MACs & 0.03 & 11.6 \\
        \hline
        \textbf{Bitonic Sorter} & 256-point & 0.24 & 486.1 \\
        \hline
        \hline
        \textbf{Total} & - & 4.11 & 620.3 \\
        \hline
    \end{tabular}
    \label{tab:hard_imp}
\end{table}

\noindent\textbf{Benchmarks.} 
All experiments are conducted on a server equipped with an AMD EPYC 7302 16-core CPU, 512 GB DDR4-3200 memory, and a NVIDIA A100 40GB GPU. The system includes an 8 TB PCIe 4.0 NVMe SSD, ensuring high-bandwidth storage access for large-scale datasets and index files.
We build IVF-PQ indexes using Faiss~\cite{johnson2019billion} with 16-byte PQ codes for all datasets. The cuVS integration~\cite{cuvs2025} is enabled within Faiss, allowing faster query search. We compare different designs using throughput (QPS) and query latency (ms) under a specific recall $@ k=100$.

\begin{table}[ht]
    \footnotesize
    \centering
    \caption{Statistics of Evaluated Datasets}
    \label{tab:datasets}
    \begin{tabular}{lrrrrr}
        \hline
        \textbf{Dataset} & \textbf{\# Vector} & \textbf{\# Dim.} & \textbf{\# Bit} & \textbf{Raw Size} & \textbf{$N_{\text{list}}$}\\
        \hline
        BIGANN-1B~\cite{jegou2011searching}     & 1B   & 128 & 8  & 119~GB & 128k \\
        SPACEV-1B~\cite{spacev1b2023}           & 1B   & 100 & 8  & 93~GB  & 128k \\
        Wiki-88M~\cite{wikiall2024}             & 88M  & 768 & 32 & 252~GB & 16k \\
        \hline
    \end{tabular}
\end{table}

\noindent\textbf{Datasets.} 
\design is evaluated on three large-scale vector datasets spanning diverse domains. BIGANN-1B~\cite{jegou2011searching} and SPACEV-1B~\cite{spacev1b2023} are two billion-scale datasets with 128-dim and 100-dim embeddings, respectively. Wiki-88M~\cite{wikiall2024} provides 88M high-dimensional (768-dim) embeddings representative of LLM retrieval workloads. Table~\ref{tab:datasets} summarizes the key statistics of these datasets.

\subsection{Performance Evaluation}

\begin{figure}[t]
    \centering
    \includegraphics[width=\linewidth]{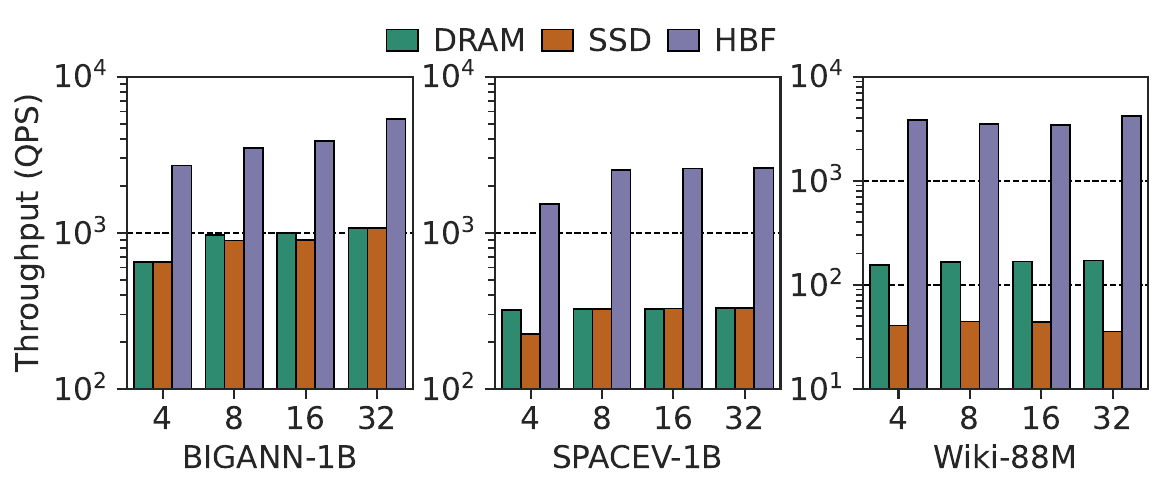}
    \caption{Throughput vs. batch size for DRAM, SSD, and HBF.}
    \label{fig:perf_qps_vs_bsz}
\end{figure}

We measure the performance by fixing recall at 0.95 for BIGANN-1B and SPACEV-1B, and at 0.9 for Wiki-88M.

\noindent\textbf{Throughput.} 
Figure~\ref{fig:perf_qps_vs_bsz} shows the throughput of DRAM, SSD, and HBF across batch sizes. HBF achieves the highest QPS on all datasets and scales smoothly with batch size, while DRAM and SSD quickly saturate due to data-movement bottlenecks. On BIGANN-1B and SPACEV-1B, HBF yields roughly 3–8$\times$ higher throughput, and its advantage grows to over 20$\times$ on the high-dimensional Wiki-88M dataset, whose reranking traffic is much larger.

\begin{figure}[t]
    \centering
    \includegraphics[width=\linewidth]{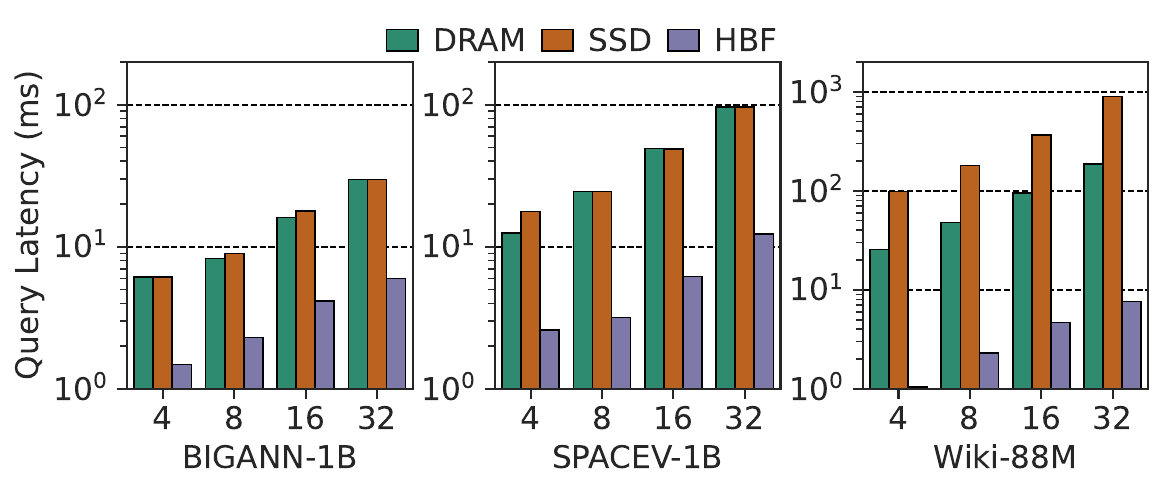}
    \caption{Query latency vs. batch size for DRAM, SSD, and HBF.}
    \label{fig:perf_lat_vs_bsz}
\end{figure}

\noindent\textbf{Latency.} 
Figure~\ref{fig:perf_lat_vs_bsz} reports query latency across batch sizes. HBF maintains the lowest latency (around 10~ms), while DRAM and especially SSD see steep increases due to off-chip reranking traffic. On BIGANN-1B and SPACEV-1B, HBF is 3–6$\times$ faster than DRAM and over an order of magnitude faster than SSD, with an even larger 10–40$\times$ advantage on the high-dimensional Wiki-88M dataset. By keeping raw vectors on-package, HBF avoids PCIe and NVMe transfers, keeping reranking latency tightly bounded.

\begin{figure}[t]
    \centering
    \includegraphics[width=\linewidth]{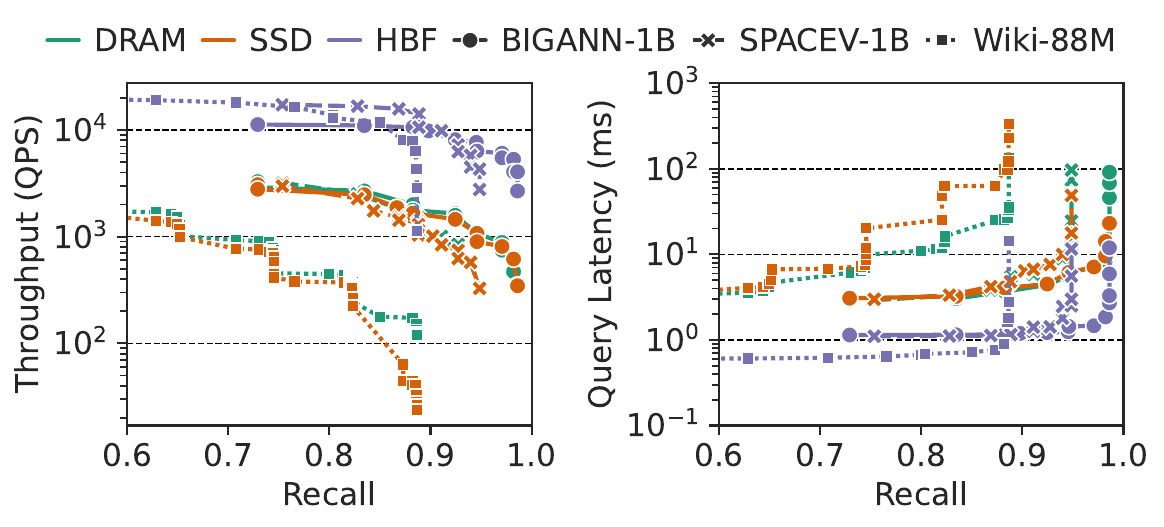}
    \caption{Throughput and recall tradeoffs.}
    \label{fig:perf_tp_recall_lat}
\end{figure}

\noindent\textbf{Throughput and Recall Tradeoffs.} 
Figure~\ref{fig:perf_tp_recall_lat} shows the throughput–recall tradeoff. HBF consistently Pareto-dominates DRAM and SSD, sustaining higher QPS at a given recall and higher recall at a given QPS. The advantage widens at high recall ($\ge0.9$), where growing reranking traffic causes sharp throughput and latency penalties for DRAM/SSD due to off-chip transfers. By keeping reranking on-package, HBF maintains both high throughput and low latency even as recall targets tighten.

\begin{figure}[t]
    \centering
    \includegraphics[width=0.75\linewidth]{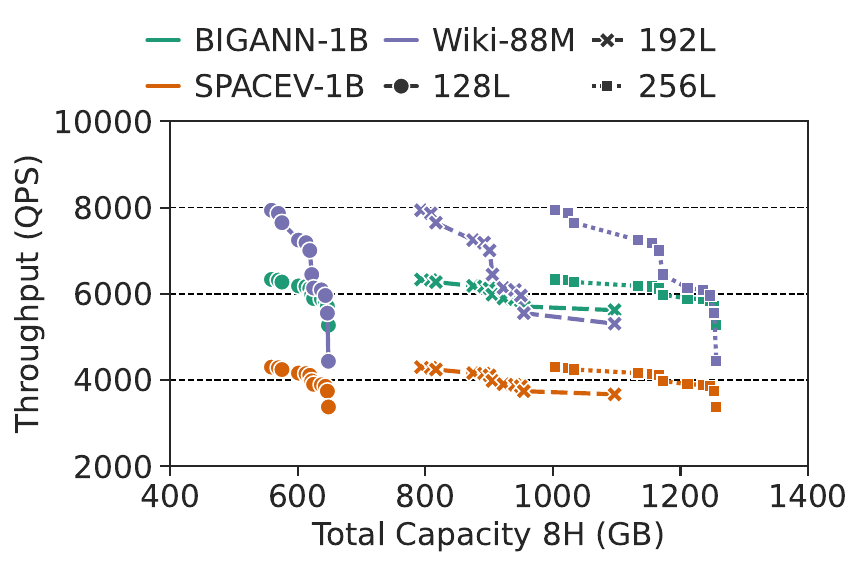}
    \caption{Throughput and capacity tradeoffs.}
    \label{fig:perf_density_qps}
\end{figure}

\noindent\textbf{Design Space Exploration.}
To validate our chosen subarray configuration (4 KB pages, 64 blocks, 256 layers), we perform a design space exploration. Figure~\ref{fig:perf_density_qps} shows the capacity–throughput tradeoff: increasing WL layers, block count, or page size improves capacity but reduces throughput. Among the 128L/192L/256L options, 256L provides the best balance, and the 4 KB–64-block configuration offers the strongest overall capacity–performance tradeoff, confirming our baseline choice.

\begin{table}[t]
    \centering
    \footnotesize
    \caption{Comparison of ANNS Accelerators on the Billion-Scale BIGANN-1B Dataset at Recall $\ge 0.9$}
    \label{tab:ann_accel_comparison}
    \setlength{\tabcolsep}{5pt}
    \renewcommand{\arraystretch}{1.1}
    \begin{tabular}{c|c|c|c}
        \hline
        \textbf{Design} & ANNA~\cite{lee_ANNASpecialized_2022} & SmartANNS~\cite{tian_ScalableBillionpoint_2024} & \design (This Work) \\
        \hline
        \hline
        \textbf{Platform}              & ASIC      & Near DRAM & GPU-HBF \\
        \hline
        \textbf{Index Algorithm}       & IVF-PQ    & Graph     & IVF-PQ  \\
        \hline
        \textbf{Dataset}               & \multicolumn{3}{c}{BIGANN-1B} \\
        \hline
        \multirow{2}{*}{\textbf{Throughput (QPS)}}      &  4.2k      &  0.9k      & 8.1k \\
                                                        &  (4.7$\times$)      &  (1.0$\times$)      & (9.0$\times$) \\
        \hline
    \end{tabular}
\end{table}

\noindent\textbf{Comparison with Other Accelerators.}
Table~\ref{tab:ann_accel_comparison} compares \design with two state-of-the-art ANNS accelerators on the billion-scale BIGANN-1B dataset at recall~$\ge 0.9$. ANNA~\cite{lee_ANNASpecialized_2022} is an IVF-PQ ASIC accelerator using 16-byte PQ codes, while SmartANNS~\cite{tian_ScalableBillionpoint_2024} is a graph-based system deployed on 12 SmartSSDs. \design achieves 8.1k QPS, delivering 1.9$\times$ and 9$\times$ speedups over ANNA and SmartANNS, respectively. By integrating on-package HBF storage with near-storage reranking, \design exploits high internal bandwidth and eliminates off-chip bottlenecks, enabling it to outperform both specialized ASIC solutions and SmartSSD-based accelerators.




\section{Conclusion}
This work presents \design, a GPU memory architecture augmented with HBF to overcome the capacity and data-movement bottlenecks of IVF-PQ reranking in large-scale ANNS. By re-architecting 3D NAND into distributed subarrays for HBF and integrating HBF within the GPU package, \design enables terabyte-scale on-package vector storage with high read bandwidth. A near-storage search unit further executes reranking inside the HBF stack, eliminating off-chip transfers. 

Our evaluation shows that \design improves reranking throughput by up to 20$\times$ and latency up to 40$\times$ across billion-scale datasets compared to GPU-DRAM and GPU-SSD systems, highlighting HBF as a promising direction for future memory-centric AI accelerators targeting retrieval-heavy workloads such as RAG.


\clearpage
\begin{footnotesize}
    \bibliographystyle{ACM-Reference-Format}
    \bibliography{refs}
\end{footnotesize}

\end{document}